\documentclass[pra,showpacs,eqsecnum,twocolumn]{revtex4}
\usepackage{amsfonts,amssymb,bm}

\begin{document}

\title{Quantum phases of a qutrit}

\author{Andrei B. Klimov}
\affiliation{ Departamento de F\'{\i}sica,
Universidad de Guadalajara,
Revoluci\'on~1500, 44420~Guadalajara,
Jalisco, Mexico}

\author{Luis L. S\'anchez-Soto}
\affiliation{Departamento de \'Optica,
Facultad de F\'{\i}sica,
Universidad Complutense,
28040 Madrid, Spain}

\author{Hubert de Guise}
\affiliation{Department of Physics,
Lakehead University, Thunder Bay,
Ontario P7B 5E1, Canada}

\author{Gunnar Bj\"{o}rk}
\affiliation{Department of Microelectronics
and Information Technology,
Royal Institute of Technology (KTH),
Electrum 229, SE-164 40 Kista, Sweden}

\date{\today}

\begin{abstract}
We consider various approaches to treat the
phases of a qutrit. Although it is possible to
represent qutrits in a convenient geometrical
manner by resorting to a generalization of the
Poincar\'e sphere, we argue that the appropriate
way of dealing with this problem is through
phase operators associated with the algebra
su(3). The rather unusual properties of these
phases are caused by the small dimension
of the system and are explored in detail.
We also examine the positive operator-valued
measures that can describe the qutrit phase
properties.
\end{abstract}


\maketitle

\section{Introduction}

The emerging field of quantum information, which
embraces areas of futuristic technology such as
quantum computing, quantum cryptography, and quantum
communications, has been built on the concepts
of entanglement and qubits~\cite{Nie00,BEZ00}.
The full appreciation of the complex quantal
properties of these two ideas has provided
powerful physical resources for new schemes
that herald results that cannot be achieved
classically~\cite{Gal02}.

Recently, the exploration of higher dimensional
quantum systems has finally received the attention
it rightly deserves. One could think that this
represents a mere digression in a hot topic.
However, qutrits have several interesting
properties worth exploring~\cite{Cav00}:
the efficiency and security of many quantum
information protocols are improved using
qutrits~\cite{Bec00,Cer02,Bru02}, and larger
violations of nonlocality via Bell tests are
expected to occur for systems of entangled
qutrits~\cite{Aci02,Col02}.

In the modern parlance of quantum information
the concept of phase for a qubit (or a qutrit)
is ubiquitous. However, this notion is rather
imprecise. Phases for three-level systems have
been handled by invoking fuzzy concepts such as
the phase of the associated wavefunction~\cite{Buc86}.
Sometimes, the problem is reduced to the
optimal estimation of the value of the phase
shift undergone by the qutrit~\cite{Dar98}.

When comparing phases of two states, it is
usually assumed that the relative phase is
obtained from the argument of their inner product.
In this perspective, the phase is considered as
a state parameter. In recent years, we have learned
that this relative phase shift can be of various
origins, namely, it can be purely dynamical or
purely geometrical or both. Presently,
there is an immense interest in geometric
phases in quantum optics~\cite{Uhl86,Sjo00},
especially in connection with quantum computing
applications~\cite{Jon00}. In fact, these phases
are linked to the geometry of the state space:
for a qubit, this space is the coset space
SU(2)/U(1), the well-known Poincar\'e sphere,
while for a qutrit, a geometrical picture of the
corresponding generalization to SU(3)/U(2) has
been recently presented~\cite{Arv97}.

We emphasize that these notions, though well
established in the classical limit, are not
easily extrapolated into the realm of the
quantum world. Since the phase is a physical
property, it must, in the orthodox picture
of quantum mechanics, be associated with a
selfadjoint operator or at least with a family
of positive operator-valued measures (POVMs).
In this spirit, phase operators for the algebra
su(2), which describes qubits, have been
previously worked out~\cite{Lev73,Vou90,Ell90,Lui97},
as well as the optimal POVM for this
problem~\cite{San95}. The main goal of this
paper is to work out a nontrivial extension
to su(3) of the results available for su(2),
enabling us to introduce phase operators for
qutrits with a clear physical picture. This
seems of such fundamental importance,
that it is surprising that such a task has
not been undertaken long time ago. We
thus trust that this will be of relevance
to workers in the various experimental fields
currently under consideration for quantum
computing technology and in quantum optics,
in general.

\section{Poincar\'e sphere for a qutrit}

We first briefly recall the salient features
of the Poincar\'e sphere representation for a
qubit, with a view of preparing its generalization
for a qutrit along the lines of Ref.~\cite{Arv97}.
A qubit lives in a two-dimensional complex
Hilbert space $\mathcal{H}^{(2)}$ spanned
by two states: $| 1 \rangle$ and $| 2 \rangle$.
To get a useful parametrization of the state space
of a general qubit described by the density matrix
$\hat{\rho}$, we observe that the Pauli
matrices $\hat{\sigma}_a$ together with the
identity $\hat{\openone}$ form a complete
set of linearly independent observables, and
that any  selfadjoint (trace class) operator
can then be written as
\begin{equation}
\hat{\rho} =  \frac{1}{2}
( \hat{\openone} +
\mathbf{n} \cdot \hat{\bm{\sigma}} ) .
\end{equation}
The physical condition $\hat{\rho} \ge 0$ holds
only when $ | \mathbf{n} | \le 1$. Hence, the state
space coincides with the Bloch ball, and the set
of pure states ($\hat{\rho}^2 = \hat{\rho}$)
with the boundary of this ball $| \mathbf{n} | = 1$,
which is the Bloch sphere $\mathcal{S}^2$.
The general pure state
\begin{equation}
| \Psi \rangle = \sin(\theta/2) | 1 \rangle
+ e^{i \phi} \cos(\theta/2) | 2 \rangle,
\end{equation}
is represented by the unit vector
\begin{equation}
\mathbf{n} = (\sin \theta \cos \phi,
\sin \theta \sin \phi,
\cos \theta) .
\end{equation}
The angle $\theta$ is obviously related to the
proportion of $| 1 \rangle$ and $| 2 \rangle$
in the composition of the state, while
the parameter $\phi$ is routinely interpreted
as the quantum phase associated with the
qubit and canonically conjugate to the inversion
$\hat{\sigma}_z$~\cite{San02}. We note in passing that
diametrically opposite points on $\mathcal{S}^2$
correspond to mutually orthogonal vectors in
$\mathcal{H}^{(2)}$.

Consider now a qutrit, living in a three-dimensional
complex Hilbert space $\mathcal{H}^{(3)}$ spanned
by $| 1 \rangle$, $| 2 \rangle$, and $| 3 \rangle$.
The roles of SU(2) and the Pauli matrices are now
played by the group SU(3) and the eight generators
of the corresponding su(3) algebra. A convenient
set of Hermitian generators are the Gell-Mann
matrices~\cite{Gel64,Wei97} $\hat{\lambda}_r$
($r= 1, \ldots, 8$), which obey the commutation
relations
\begin{equation}
[\hat{\lambda}_r, \hat{\lambda}_s ] = 2
i f_{rst} \hat{\lambda}_t ,
\end{equation}
where, above and in the following, the summation
over repeated indices applies. The structure constants
$f_{rst}$ are elements of a completely antisymmetric
tensor spelled out explicitly in Ref.~\cite{Arv97},
for example.

A particular feature of the generators of SU(3)
in the defining $3 \times 3$ matrix representation
is closure under anticommutation~\cite{Wei97}
\begin{equation}
\{ \hat{\lambda}_r, \hat{\lambda}_s \} =
\frac{4}{3} \delta_{rs} \hat{\openone}
+ 2  d_{rst} \hat{\lambda}_t ,
\end{equation}
where now $d_{rst}$ form a totally symmetric
tensor.

For the following, a vector-type notation is
useful, based on the structure constants. The
$f$ and $d$ symbols allow us to define both
antisymmetric and symmetric products by
\begin{eqnarray}
(\mathbf{A} \wedge \mathbf{B})_r & = &
f_{rst} A_s B_t =
-(\mathbf{B} \wedge \mathbf{A})_r \nonumber \\
& & \\
(\mathbf{A} \star \mathbf{B})_r & = &
\sqrt{3} d_{rst} A_s B_t =
+ ( \mathbf{B} \star \mathbf{A} )_r .
\nonumber
\end{eqnarray}

Given a density matrix $\hat{\rho}$ we can
expand it in terms of the unit matrix
$\hat{\openone}$ and the $\hat{\lambda}_{r}$
in the form
\begin{equation}
\hat{\rho} = \frac{1}{3}(1+ \sqrt{3} \mathbf{n}
\cdot \hat{\bm{\lambda}} ) .
\end{equation}
This is the equivalent to the Bloch ball for
a qutrit. For a pure state the analogous Bloch
sphere is defined by the condition
\begin{equation}
\mathbf{n} \cdot \mathbf{n}=1,
\qquad
\mathbf{n} \star \mathbf{n}= \mathbf{n} .
\end{equation}
Thus, each pure qutrit state corresponds
to a unique unit vector $\mathbf{n} \in \mathcal{S}^7$,
the seven-dimensional unit sphere. In addition,
this vector must obey the condition $\mathbf{n}
\star \mathbf{n} = \mathbf{n}$, which places three
additional constraints, thus reducing the number of
real parameters required to specify a pure state
from the seven parameters needed to specify an
arbitrary eight-dimensional vector to four.

In view of our discussion for qubits, it is clear
that normalization and a choice for the arbitrary
overall phase allow us to write these four parameters
for any pure state as
\begin{eqnarray}
\label{puqu}
| \Psi \rangle & = &  \sin (\xi/2)
\cos (\theta/2)  | 1 \rangle \nonumber \\
& + &   e^{i \phi_{12}}
\sin (\xi/2) \sin (\theta/2) | 2 \rangle
\nonumber \\
& + & e^{i \phi_{13}} \cos (\xi/2) | 3 \rangle ,
\end{eqnarray}
Again, $\theta$ and $\xi$ determine
the magnitudes of the components of $| \Psi \rangle$,
while we can interpret $\phi_{12}$ as the phase of
$| 1 \rangle$ relative to $| 2 \rangle$ and analogously
for $\phi_{13}$.  We can easily obtain the expressions
for $\mathbf{n}$ in these local coordinates.

Some interesting geometric properties of this
Poincar\'e sphere are discussed in Ref.~\cite{Cav00}.
In particular, it is easily seen that for
two unit vectors $\mathbf{n}$ and $\mathbf{n}^\prime$
representing pure states
\begin{equation}
0 \leq \arccos (\mathbf{n}\cdot
\mathbf{n}^{\prime} ) \leq \frac{2 \pi}{3} ,
\end{equation}
so mutually orthogonal vectors in $\mathcal{H}^{(3)}$
do not lead to antipodal or diametrically opposite
points on the Poincar\'e sphere, but to points
with a maximum opening angle of $2 \pi/3$.

\section{Phase operators for a qutrit}

Although there is a widespread usage of dealing
with the qutrit phases as state parameters,
this is not an orthodox way of proceeding,
according to the very basic principles of
quantum mechanics.

To gain further insights into this obvious
although almost unnoticed point, we stress
that the complete description of a
qutrit involves the nine operators
\begin{equation}
\label{S}
\hat{S}_{ij} = | i \rangle \langle j | ,
\end{equation}
where $| i \rangle$ is a basis vector in
$\mathcal{H}^{(3)}$. The three ``diagonal"
operators $\hat{S}_{ii}$ measure level
populations, while the ``off-diagonal''
ones $\hat{S}_{ij}$ represent transitions from
$j$ to level $i$. One can easily check
that they satisfy
\begin{equation}
\label{ccr3}
[\hat{S}_{ij}, \hat{S}_{kl} ] =
\delta_{jk} \hat{S}_{il}
- \delta_{il} \hat{S}_{kj} ,
\end{equation}
which are the commutation relations of the
algebra u(3)~\cite{Gil74}.

Because of the trivial constraint $\hat{S}_{11}
+ \hat{S}_{22} + \hat{S}_{33}= \hat{\openone}$, only
two populations can vary independently. For this
reason, we shall work with two independent
traceless operators
\begin{equation}
\hat{S}_{12}^z = \frac{1}{2}
(\hat{S}_{22} - \hat{S}_{11}) ,
\qquad
\hat{S}_{23}^z = \frac{1}{2}
(\hat{S}_{33} - \hat{S}_{22}) ,
\end{equation}
that measure atomic inversions between the
corresponding levels. In atomic systems,
the selection rules usually rule out one
of the transitions and therefore the
two independent inversions are automatically
fixed. For a general qutrit, these inversions
can be arbitrarily chosen.

The commuting operators $\hat{S}_{12}^z$ and
$\hat{S}_{23}^z$ constitute a maximal abelian
subalgebra for the qutrit (known as Cartan subalgebra).
From the discussion of the previous section,
we expect $\hat{S}_{12}^z$ and $\hat{S}_{23}^z$
to be conjugate to the corresponding
(independent) phases of the qutrit.

Note that $(\hat{S}_{12}, \hat{S}_{12}^z)$ and
$(\hat{S}_{23}, \hat{S}_{23}^z)$ correspond to
the qubits $1 \leftrightarrow 2$ and
$2 \leftrightarrow 3$. However, these two
qubits are not independent, since Eq.~(\ref{ccr3})
imposes highly nontrivial coupling between
them.

At the operator level, the equivalent to
the decomposition of a complex number in
terms of modulus and phase is a polar
decomposition~\cite{Lui93}. Since $\hat{S}_{21}
= \hat{S}^\dagger_{12}$, it seems appropriate
to define~\cite{Kli03}
\begin{equation}
\label{polar}
\hat{S}_{12} =
\hat{R}_{12}  \ \hat{E}_{12} ,
\end{equation}
where the ``modulus" is $\hat{R}_{12} =
\sqrt{\hat{S}_{12} \hat{S}_{21}}$ and $\hat{E}_{12}=
\exp(i \hat{\phi}_{12})$, $\hat{\phi}_{12}$ being the
Hermitian operator representing the phase.

One can easily work out that a unitary solution of
Eq.~(\ref{polar}) is given, up to an overall phase,
by
\begin{equation}
\hat{E}_{12} = | 1 \rangle \langle 2 |
+ e^{i \phi_0} \ | 2 \rangle \langle 1 |
- e^{-i \phi_0} \ | 3 \rangle \langle 3 | ,
\end{equation}
where the undefined factor $e^{ i \phi_0}$
appears due to the unitarity requirement of
$\hat{E}_{12}$. The main features of
this operator are largely independent of
$\phi_0$, but for the sake of concreteness,
we can make a definite choice. For
example~\cite{Lui97}, for a qubit
defined by a linear superposition of
the states $| 1 \rangle$ and $| 2 \rangle$,
the complex conjugation of the wave function
should reverse the sign of $\hat{\phi}_{12}$,
which immediately leads to the condition
$e^{i \phi_0}= -1$. We conclude then that
a unitary phase operator that preserves
the polar decomposition of Eq.~(\ref{polar})
can be represented as
\begin{equation}
\hat{E}_{12} = | 1 \rangle \langle 2 |
- | 2 \rangle \langle 1 |
+  | 3 \rangle \langle 3 | .
\end{equation}
The eigenstates of $\hat{\phi}_{12}$ are
those of $\hat{E}_{12}$, and easily found to
be
\begin{equation}
| \phi_{12}^0 \rangle  =  | 3 \rangle ,
\qquad
| \phi_{12}^\pm \rangle  =
\frac{1}{\sqrt{2}}
( | 2 \rangle \pm i | 1 \rangle ) ,
\end{equation}
with the corresponding eigenvalues of
$\hat{\phi}_{12}$ 0 and $\mp \pi/2$,
respectively. This is a remarkable result.
It shows that the eigenvectors
$| \phi_{12}^\pm \rangle$ look like
the standard ones for a qubit. However, the
``spectator'' level $| 3 \rangle$ is an
eigenstate of this operator, which introduces
drastic changes. In other words, the phase of
the qubit $1 \leftrightarrow 2$ ``feels" the state
$| 3 \rangle$.

An analogous reasoning for the transition
$2 \leftrightarrow 3$ gives the corresponding
operator $\hat{E}_{23}$
\begin{equation}
\hat{E}_{23} = | 2 \rangle \langle 3 |
- | 3 \rangle \langle 2 |
+  | 1 \rangle \langle 1 | ,
\end{equation}
with eigenvectors
\begin{equation}
| \phi_{23}^0 \rangle  =  | 1 \rangle ,
\qquad
| \phi_{23}^\pm \rangle  =
\frac{1}{\sqrt{2}}
( | 3 \rangle \pm i | 2 \rangle ) ,
\end{equation}
and the same spectrum as before.

As for the operator $E_{13}$, one must
be careful, because it connects the lowest
to the highest vector. In fact, the polar
decomposition in this case gives as
a unitary solution
\begin{equation}
\label{gen13}
\hat{E}_{13} = a | 3 \rangle \langle 2 |
- b^\ast | 3 \rangle \langle 1 | +
b  | 2 \rangle \langle 2 |+ a^\ast | 2 \rangle
\langle 1 | + | 1 \rangle \langle 3 | ,
\end{equation}
with the condition $| a |^2 + | b |^2 = 1.$
There are also nonunitary solutions to the
polar decomposition, but they lack of interest
to describe a phase observable in our context.

Note that the general solution (\ref{gen13}) has
the desirable property $\hat{E}_{13} |3 \rangle =
| 1 \rangle$. On physical grounds, we argue that
the state $| 2 \rangle $ should be a ``spectator"
for the transition $1 \leftrightarrow 3$.
Thus we impose $\hat{E}_{13} |2 \rangle \propto
| 2 \rangle$, which is only possible if $a = 0$
and we have that
\begin{equation}
\hat{E}_{13} = | 1 \rangle \langle 3 |
- | 3 \rangle \langle 1 |
+  | 2 \rangle \langle 2 | ,
\end{equation}
with eigenvectors
\begin{equation}
| \phi_{13}^0 \rangle  =  | 2 \rangle ,
\qquad
| \phi_{13}^\pm \rangle =
\frac{1}{\sqrt{2}}
( | 3 \rangle \pm i | 1 \rangle ) .
\end{equation}
With this choice we are led to
\begin{equation}
\label{noncom}
\hat{E}_{12} \hat{E}_{23}   \neq
\hat{E}_{13} ,
\end{equation}
which clearly displays the quantum nature of this
phase~\cite{Gui02}. Note, in passing, that
\begin{equation}
[\hat{E}_{12}, \hat{R}_{23} ] =
[\hat{E}_{23}, \hat{R}_{12}] = 0 ,
\end{equation}
and $[\hat{R}_{23}, \hat{R}_{12}] = 0$, so the
interference between different channels (i.e., the
noncommutativity of $\hat{S}_{12}$ and $\hat{S}_{23}$)
is due to the nonncommutativity of the corresponding phases.

\section{Positive operator measures
for the qutrit phases}

The unusual behavior exhibited by the
description of qutrit phases in terms of
Hermitian operators can be considered to
some extent exotic. One may think it
preferable to represent qutrit phases
by using a positive operator-valued measure
(POVM) taking continuous values in a
$2 \pi$ interval.

We briefly recall that a POVM~\cite{Hel76}
associated to an observable $\hat{\phi}$ is
a set of linear operators $\hat{\Delta} (\phi)$
($0 \le \phi < 2 \pi$), depending on the
continuous parameter $\phi$ and furnishing
the correct probabilities in any measurement
process through the fundamental postulate that
\begin{equation}
P(\phi) = \mathrm{Tr} [
\hat{\rho} \ \hat{\Delta} (\phi) ].
\end{equation}
The real valuedness, positivity, and
normalization of $P(\phi)$ impose
\begin{equation}
\label{condpom}
\hat{\Delta}^\dagger (\phi) = \hat{\Delta} (\phi) ,
\quad
\hat{\Delta} (\phi) \ge 0 ,
\quad
\int_0^{2 \pi} d\phi \ \hat{\Delta} (\phi) =
\hat{\openone} .
\end{equation}
where the integral extends over any $2 \pi$
interval of the form $(\phi_0, \phi_0 + 2 \pi)$,
$\phi_0$
being a fiducial or reference phase. Note that,
in general, $\hat{\Delta} (\phi)$ are not
orthogonal projectors as in the standard
von Neumann measurements described by
selfadjoint operators.

From our previous discussion, it is clear
that we expect some complementarity between phases
and inversions~\cite{Leo95,Bus95,Lui98}.  If we
observe that
\begin{equation}
e^{ i  \phi^\prime \hat{S}_{12}^z} =
e^{ -i  \phi^\prime/2} | 1 \rangle \langle 1 |
+ e^{ i  \phi^\prime/2} | 2 \rangle \langle 2 | +
| 3 \rangle \langle 3 | ,
\end{equation}
and argue that phase-shift operators must be
$2 \pi$ periodic, we impose that any POVM
$\hat{\Delta} (\phi_{12}, \phi_{23})$ for a
qutrit should satisfy
\begin{eqnarray}
\label{req1}
e^{ i 2 \phi^\prime \hat{S}_{12}^z}
\hat{\Delta} (\phi_{12}, \phi_{23})
e^{ - i 2  \phi^\prime \hat{S}_{12}^z}
& = & \hat{\Delta} (\phi_{12}+ \phi^\prime,
\phi_{23} ) , \nonumber \\
& & \\
e^{ i 2 \phi^{\prime \prime} \hat{S}_{23}^z}
\hat{\Delta} (\phi_{12}, \phi_{23})
e^{ - i 2 \phi^{\prime \prime} \hat{S}_{23}^z} & = &
\hat{\Delta} (\phi_{12}, \phi_{23} + \phi^{\prime \prime} ) .
\nonumber
\end{eqnarray}
One can work out that the general POVM  fulfilling
these requirements must be of the form
\begin{eqnarray}
\label{Dgen}
\hat{\Delta} (\phi_{12}, \phi_{23})
& = & \frac{1}{(2 \pi)^2} \{ \hat{\openone} +
[ \gamma_{12} e^{i (2 \phi_{12} - \phi_{23})}
| 2 \rangle \langle 1 | \nonumber \\
& + &  \gamma_{23} e^{i (2 \phi_{23}- \phi_{12}) }
| 3 \rangle \langle 2 | +  \gamma_{13}
e^{i (\phi_{12}+ \phi_{23})} | 3 \rangle \langle 1 |
\nonumber \\
& + & \mbox{\textrm{h. c.}} ] \} ,
\end{eqnarray}
where h. c. denotes Hermitian conjugate,
$\gamma_{ij} \le 1$ are real numbers and
$\phi_{ij}$ is the relative phase between
states $| i \rangle$ and $| j \rangle$.
These relative phases coincide precisely with
the polar part of the realization of su(3) on
the torus constructed in Ref.~\cite{Gui02}.
If we chose the $\gamma_{ij}$ different, say
$\gamma_{12} =1$ and the other two below the
unity, then the expectation value of this
POVM could reach the value zero for the superposition
states $(|1 \rangle + \exp(i \theta) | 2 \rangle) /
\sqrt{2}$. However, for superpositions of
states $|1 \rangle$ and $|3 \rangle$ or
 $|2 \rangle$ and $|3 \rangle$, the expectation
values of the POVM would always be greater than
zero. Since there is no physical reason to
assign special relevance to one specific
superposition of the states, we assume that
the POVM must be symmetric with respect to
the states, which leads to
\begin{equation}
\gamma \equiv \gamma_{12} = \gamma_{23} =
\gamma_{13} .
\end{equation}
Moreover, we make henceforth the choice $\gamma = 1$
because only for this choice the POVM can attain
the expectation value zero for some particular state.

In contrast with the result of Eq.~(\ref{noncom})
formulated in terms of operators, now
there are only two relevant phases in the qutrit
description: the third can be inferred from the
other two, as in the classical description.

The proposed POVM provides qutrit phases
where any values of $\phi_{12}$
and $\phi_{23}$ are allowed. However,
note that the probability density
induced by this POVM can be written as
\begin{eqnarray}
P(\phi_{12}, \phi_{23}) & = &
\frac{1}{(2 \pi)^2} \{ 1 + [ \rho_{12}
e^{i (2 \phi_{12} - \phi_{23})}
+
\rho_{23} e^{i (2 \phi_{23}- \phi_{12}) }
\nonumber \\
& + &  \rho_{13} e^{i (\phi_{12}+ \phi_{23})}
+ \mbox{\textrm{c. c.}} ] \}  \, ,
\end{eqnarray}
where $\rho_{ij} = \langle i | \hat{\rho} | j \rangle$
and c. c. denotes complex conjugate. Therefore, this
continuous range of variation is not effective in the
sense that the values of $P(\phi_{12}, \phi_{23})$ at
every point $(\phi_{12}, \phi_{23})$ cannot be
independent, and we can find relations between
them irrespective of the qutrit state. In other
words, the complex parameters $c_{ij}$ can
be determined by the values of $P(\phi_{12}, \phi_{23})$
at six points. Discreteness is inevitably at the heart of
the qutrit phase~\cite{San02}.

Finally, we shall consider a remarkable example
of POVM particularly suited to describe the
qutrit phase. We recall that for a single-mode
quantum field, a POVM for the field phase
can be defined in terms of radial
integration of quasiprobability distributions
obtained using a coherent-state representation,
much in the spirit of the classical
conception~\cite{Lui97b,Sch01}. The natural
generalization of this procedure to the qutrit
problem involves the use of su(3) coherent
states. In the Appendix we summarize the essential
ingredients needed for this paper. Coherent
states of a single qutrit are of the form
\begin{equation}
| \alpha, \beta \rangle = \frac{1}
{\sqrt{\mathcal{C}_{\alpha \beta}}}
( | 3 \rangle + \alpha | 2 \rangle + \alpha \beta | 1 \rangle ),
\end{equation}
where $\alpha$ and $\beta$ are complex numbers
and the normalization constant is
\begin{equation}
\mathcal{C}_{\alpha \beta} = 1 +
|\alpha|^2 (1 + |\beta |^2 ) .
\end{equation}
These coherent states generate a POVM over the
qutrit state space via the projectors
$| \alpha , \beta \rangle \langle \alpha, \beta |$.

As shown in the Appendix, the phases of
$\alpha$ and $\beta$ are just those of
$\langle \alpha , \beta | \hat{S}_{32}
| \alpha , \beta \rangle$ and $\langle \alpha , \beta |
\hat{S}_{21} | \alpha , \beta \rangle$, respectively,
while the phase associated to  $\langle \alpha , \beta |
\hat{S}_{31} | \alpha , \beta \rangle$ is just
the product of the other two. Let us write
\begin{equation}
\alpha = r_{23} e^{i \phi_{23}} ,
\qquad
\beta = r_{12} e^{i \phi_{12}} ,
\end{equation}
and integrate the projectors
$| \alpha , \beta \rangle \langle \alpha, \beta |$
radially over $r_{12}$ and $r_{23}$,  with respect
the measure [see Eq. (\ref{meas})]
\begin{equation}
d \mu =
\frac{ |\alpha|^2}{[1 + |\alpha|^2 ( 1 + |\beta |^2)]^3}
d^2 \alpha d^2 \beta .
\end{equation}
After some calculations one obtains
\begin{eqnarray}
\hat{\Delta} (\phi_{12}, \phi_{23})
& = & \frac{1}{(2 \pi)^2} \{ \hat{\openone} +
\frac{\pi}{96} [   e^{i (2 \phi_{12} - \phi_{23})}
| 2 \rangle \langle 1 | \nonumber \\
& + &   e^{i (2 \phi_{23}- \phi_{12}) }
| 3 \rangle \langle 2 |
\nonumber \\
& + & e^{i (\phi_{12}+ \phi_{23})}
| 3 \rangle \langle 1 | + \mathrm{h. c.} ] \} ,
\end{eqnarray}
which is just a specilized form of
Eq.~(\ref{Dgen}) and whose physical
meaning is now clear.

\section{Concluding remarks}

In this paper we have looked for possible
descriptions of qutrit phases. Although it
is possible to construct an extension of the
Poincar\'e sphere to qutrits, the orthodox
way of dealing with any observable is
to represent them by selfadjoint operators.
In this spirit, we have investigated a
description of qutrit phases in terms of
a proper polar decomposition of its amplitudes.
Perhaps the most striking consequence of
this description is that phases are discrete
and do not commute.

We have also considered alternative generalized
descriptions in terms of POVMs. In these
descriptions, phases appear as parameters
rather than operators. Additivity of phases
follows from the commutativity of the
Cartan elements. Although these POVMs
reflect some desirable properties of the
classical phase, they show an effective
discreteness, even if in principle a continuous
range of variation is assumed.

\begin{acknowledgments}
We wish to acknowledge Hoshang Heydari for pointing
us the correct form of the POVM associated with
a qutrit and for very helpful discussions.
The work of Hubert de Guise is supported by  NSERC of Canada,
while the work of Gunnar Bj\"ork is supported by the Swedish
Research Council (VR).
\end{acknowledgments}

\appendix*

\section{su(3) coherent states}

In this Appendix, we briefly summarize the essential
ingredients of the construction of coherent states
for three-level systems~\cite{Cha95}.
For concreteness, we shall consider fully
symmetrical states of $N$ three-level
systems. In the Fock representation, we denote by
$|n_1, n_2, n_3 \rangle$ the state in which there
are $n_1$ systems in level 1, $n_2$ systems in level 2
and $n_3$ systems in level 3. We observe that all
these states can be generated from $|0, 0, N \rangle$
by repeated application of the usual collective operators
$\hat{S}_{23}$ and $\hat{S}_{12}$ [note that they coincide
with (\ref{S}), introduced for one qutrit, when $N=1$] as
\begin{eqnarray}
 (\hat{S}_{12})^n \ (\hat{S}_{23})^m
|0, 0,  N \rangle  & = & \sqrt{\frac{N! m!}{(N-m)!}}
\sqrt{\frac{m! n!}{(m - n)!}} \nonumber \\
& \times & |n, m-n, N - m \rangle
\end{eqnarray}
with $0 \le n \le m \le N$. Note that this
is a simple extension of the relevant formula
for the two-level case. In analogy with the
atomic coherent states for su(2), we
define coherent states for qutrits as
\begin{equation}
| \alpha , \beta \rangle = \sqrt{\mathcal{N}_{\alpha \beta}}
e^{ \beta \hat{S}_{12}} \ e^{ \alpha \hat{S}_{23}}
|0, 0, N \rangle ,
\end{equation}
where $\alpha$ and $\beta$ are complex numbers and
$\mathcal{N}_{\alpha \beta}$ is a normalization constant
that we shall write as
\begin{equation}
\mathcal{N}_{\alpha \beta} = \frac{1}
{(\mathcal{C}_{\alpha \beta})^N} ,
\end{equation}
where we have introduced the real quantity
\begin{equation}
\mathcal{C}_{\alpha \beta} = 1 + |\alpha|^2 (1 + |\beta |^2 ) .
\end{equation}
In the Fock basis these states can be recast as
\begin{eqnarray}
| \alpha , \beta \rangle & = & \sqrt{\mathcal{N}_{\alpha \beta}}
\sum_{0 \le n \le m \le N}
\left (
\begin{array}{c}
N \\
m
\end{array}
\right )^{1/2}
\left (
\begin{array}{c}
m \\
n
\end{array}
\right )^{1/2} \nonumber \\
& \times &
\alpha^m \beta^n |n, m-n, N - m \rangle .
\end{eqnarray}
After some calculations one gets
the following mean values
\begin{eqnarray}
\bar{n}_3 & = & \langle \alpha , \beta |
\hat{S}_{33} | \alpha , \beta \rangle  =
\frac{N}{\mathcal{C}_{\alpha \beta}} ,
\nonumber \\
\bar{n}_2 & = & \langle \alpha , \beta |
\hat{S}_{22} | \alpha , \beta \rangle  =
\frac{N }{\mathcal{C}_{\alpha \beta}} |\alpha |^2 ,  \\
\bar{n}_1 & = & \langle \alpha \beta |
\hat{S}_{11} | \alpha , \beta \rangle  =
\frac{N } {\mathcal{C}_{\alpha \beta}}
|\alpha |^2 |\beta |^2  , \nonumber
\end{eqnarray}
which immediately shows that the ratios of
the average population numbers are given
by
\begin{equation}
\bar{n}_3 : \bar{n}_2 : \bar{n}_1 =
1 : |\alpha |^2 : |\alpha |^2 |\beta|^2 .
\end{equation}
On the other hand, one can also compute
\begin{eqnarray}
\langle \alpha , \beta | \hat{S}_{32}
| \alpha , \beta \rangle  & = &
\frac{N }{\mathcal{C}_{\alpha \beta}} \alpha ,
\nonumber \\
\langle \alpha , \beta | \hat{S}_{21}
| \alpha , \beta \rangle  & = &
\frac{N}{\mathcal{C}_{\alpha \beta}}
|\alpha |^2 \beta ,  \\
\langle \alpha , \beta | \hat{S}_{31}
| \alpha , \beta \rangle  & = &
\frac{N}{\mathcal{C}_{\alpha \beta}}  \alpha \beta .
\nonumber
\end{eqnarray}
The phases of $\alpha$ and $\beta$ are then
just those of $\langle \alpha , \beta | \hat{S}_{32}
| \alpha , \beta \rangle$ and $\langle \alpha , \beta |
\hat{S}_{21} | \alpha , \beta \rangle$, respectively.
Note, in passing, that the third phase associated
to  $\langle \alpha , \beta |
\hat{S}_{31} | \alpha , \beta \rangle$ is just
the product of the other two, as it happens in
classical physics.

The atomic coherent states with different amplitudes
are not orthogonal
\begin{eqnarray}
& & \langle \alpha_1 , \beta_1 |\alpha_2 , \beta_2 \rangle =
\nonumber \\
& & \\
& &
\frac{[1 + \alpha_1^\ast \alpha_2 ( 1 + \beta_1^\ast
\beta_2)]^N}{[1 + |\alpha_1|^2 (1 + |\beta_1|^2)]^{N/2}
[1 + |\alpha_2|^2 (1 + |\beta_2|^2)]^{N/2}} , \nonumber
\end{eqnarray}
but form an overcomplete set. In fact, it is easy to
verify the following
resolution of the identity
\begin{equation}
\frac{(N+1)(N+2)}{\pi^2} \int d\mu \
| \alpha , \beta \rangle \langle \alpha , \beta |=
\hat{\openone} ,
\end{equation}
where the measure $d\mu$ is
\begin{equation}
\label{meas}
d\mu =
\frac{ |\alpha|^2}{[1 + |\alpha|^2 ( 1 + |\beta |^2)]^3}
d^2 \alpha d^2 \beta .
\end{equation}
The above discussion pertains only to the fully
symmetric subspace. Of course, it is enough
for our purposes, although is can be extended
to other subspaces in a very direct way.


\begin{thebibliography}{99}

\bibitem{Nie00}
M. A. Nielsen and I.L. Chuang,
\textit{Quantum Computation and Quantum Information}
(Cambridge University Press, Cambridge, UK, 2000).

\bibitem{BEZ00}
\textit{The physics of quantum information},
edited by D. Bouwmeester, A. Ekert and A. Zeilinger
(Springer Verlag, Berlin, 2000).

\bibitem{Gal02}
A. Galindo and M. A. Mart\'{\i}n-Delgado,
Rev. Mod. Phys. \textbf{74}, 347 (2002)

\bibitem{Cav00}
C. M. Caves and G. J. Milburn,
Opt. Commun.\textbf{179}, 439 (2000).

\bibitem{Bec00}
H. Bechmann-Pasquinucci and A. Peres,
Phys. Rev. Lett. \textbf{85}, 3313 (2000).

\bibitem{Cer02}
N. J. Cerf, M. Bourennane, A. Karlsson, and N. Gisin,
Phys. Rev. Lett. \textbf{88}, 127902 (2002).

\bibitem{Bru02}
\v{C}. Brukner, M. \.{Z}ukowski, and A. Zeilinger,
Phys. Rev. Lett. \textbf{89}, 197901 (2002).

\bibitem{Aci02}
A. Ac\'{\i}n, T. Durt, N. Gisin, and J. I. Latorre,
Phys. Rev. A \textbf{65}, 052325 (2002).

\bibitem{Col02}
D. Collins, N. Gisin, N. Linden, S. Massar,
and S. Popescu,
Phys. Rev. Lett. \textbf{88}, 040404 (2002).

\bibitem{Buc86}
S. J. Buckle, S. M. Barnett, P. L. Knight,
M. A. Lauder, and D. T. Pegg,
J. Mod. Opt. \textbf{33}, 119 (1986).

\bibitem{Dar98}
G. M. D' Ariano, C. Macchiavello, and M. F. Sacchi,
Phys. Lett. A \textbf{248}, 103 (1998);
C. Macchiavello,
Phys. Rev. A \textbf{67}, 062302 (2003).

\bibitem{Uhl86}
A. Uhlmann, Rep. Math. Phys. \textbf{24}, 229 (1986);
Lett. Math. Phys. \textbf{21}, 229 (1991).

\bibitem{Sjo00}
E. Sj\"{o}qvist, A. K. Pati, A. Ekert, J. S. Anandan,
M. Ericsson, D. K. L. Oi, and V. Vedral,
Phys. Rev. Lett. \textbf{85}, 2845 (2000).

\bibitem{Jon00}
J. A. Jones, V. Vedral, A. Ekert, and
G. Castagnoli,
Nature (London) \textbf{403}, 869 (2000).

\bibitem{Arv97}
Arvind, K. S. Mallesh, and N. Mukunda,
J. Phys. A \textbf{30}, 2417 (1997).

\bibitem{Lev73}
J. M. L\'evy-Leblond,
Rev. Mex. Fis. \textbf{22}, 15 (1973).

\bibitem{Vou90}
A. Vourdas,
Phys. Rev. A \textbf{41}, 1653 (1990).

\bibitem{Ell90}
D. Ellinas,
J. Math. Phys. \textbf{32}, 135 (1990).

\bibitem{Lui97}
A. Luis and L. L. S\'anchez-Soto,
Phys. Rev. A \textbf{56}, 994 (1997);
L. L. S\'anchez-Soto and A. Luis,
Opt. Commun. \textbf{133}, 159 (1997).

\bibitem{San95}
B. C. Sanders and G. J. Milburn,
Phys. Rev. Lett. \textbf{75}, 2944 (1995).

\bibitem{San02}
L. L. S\'anchez-Soto, J. Delgado,
A. B. Klimov, and G. Bj\"{o}rk,
Phys. Rev. A \textbf{66}, 042112 (2002).

\bibitem{Gel64}
M. Gell-Mann and Y. Neeman,
\textit{ The Eightfold Way}
(Benjamin, New York, 1964).

\bibitem{Wei97}
S. Weigert,
J. Phys. A \textbf{30}, 8739 (1997).

\bibitem{Gil74}
R. Gilmore,
\textit{Lie Groups, Lie Algebras, and Some of Their Applications}
(Wiley, New York, 1974).

\bibitem{Lui93}
A. Luis and L. L. S\'anchez-Soto,
Phys. Rev. A \textbf{48}, 4702 (1993).

\bibitem{Kli03}
A. B. Klimov, L. L. S\'anchez-Soto,
J. Delgado, and E. C. Yustas,
Phys. Rev. A \textbf{67}, 013803 (2003).

\bibitem{Gui02}
H. de Guise and M. Bertola,
J. Math. Phys. \textbf{43}, 3425 (2002).

\bibitem{Hel76}
C. W. Helstrom,
\textit{Quantum Detection and Estimation Theory}
(Academic, New York, 1976).

\bibitem{Leo95}
U. Leonhardt, J. A. Vaccaro,
B. B\"{o}hmer, and H. Paul,
Phys. Rev. A \textbf{51}, 84 (1995).

\bibitem{Bus95}
P. Busch, M. Grabowski, and P. Lahti,
Ann. Phys. (N.Y.) \textbf{237}, 1 (1995).

\bibitem{Lui98}
A. Luis and L. L. S\'anchez-Soto,
Eur. Phys. J. D \textbf{3}, 195 (1998).

\bibitem{Lah99}
P. Lahti and J. P. Pellonp\"{a}\"{a},
J. Math. Phys. \textbf{40}, 4688 (1999);
\textit{ibid} \textbf{41}, 7352 (2000).

\bibitem{Lui97b}
A. Luis and L. L. S\'anchez-Soto,
Phys. Rev. A \textbf{48}, 752 (1993).

\bibitem{Sch01}
W. P. Schleich,
\textit{Quantum Optics in Phase Space}
(Wiley, Weinheim, 2001).

\bibitem{Cha95}
C. Chang-qi and F. Haake,
Phys. Rev. A \textbf{51}, 4203 (1995).
\end{thebibliography}
\end{document}